\documentclass[aps,pr,twocolumn]{revtex4-1}

\usepackage{graphicx,amsmath}

\usepackage{color}
\usepackage[unicode=true,colorlinks=true,citecolor=blue,urlcolor=blue]{hyperref}

\newcommand{\e}{{\rm e}}
\renewcommand{\i}{{\rm i}}
\renewcommand{\d}{\mathrm d}

\begin{document}

\title{Nuclear magnetic resonance and nuclear spin relaxation in AlAs quantum well\\ probed by ESR}

\author{A.~V.~Shchepetilnikov$^{1}$\email{shchepetilnikov@issp.ac.ru}, D.~D.~Frolov$^{1}$, Yu.~A.~Nefyodov$^{1}$,
I.~V.~Kukushkin$^{1}$, D.~S.~Smirnov$^{2}$, L.~Tiemann$^{3}$,
C.~Reichl$^{3}$, W.~Dietsche$^{3}$, and W.~Wegscheider$^{3}$}

\affiliation{$^{1}$Institute of Solid State Physics RAS, 142432
\protect{Chernogolovka, Moscow district, Russia}\\
{%$^2$Ioffe Institute, 194021 St.~Petersburg, Russia\\
 ~~~$^2$Ioffe Institute, 194021 St.~Petersburg, Russia~~~\\}
$^{3}$Solid State Physics
Laboratory, ETH Zurich, Schafmattstrasse 16, 8093 Zurich,
Switzerland}

\begin{abstract}

The study of nuclear magnetic resonance and nuclear spin-lattice
relaxation was conducted in an asymmetrically doped to
$n\sim1.8\times10^{11}$~cm$^{-2}$ 16~nm AlAs quantum well grown in
the $[001]$-direction. Dynamic polarization of nuclear spins due to
the hyperfine interaction resulted in the so-called Overhauser shift
of the two-dimensional conduction electron spin resonance. The
maximum shifts achieved in the experiments are several orders of
magnitude smaller than in GaAs-based heterostructures indicating
that hyperfine interaction is weak. The nuclear spin-lattice
relaxation time extracted from the decay of Overhauser shift over
time turned out to depend on the filling factor of the
two-dimensional electron system. This observation indicates that
nuclear spin-lattice relaxation is mostly due to the interaction
between electron and nuclear spins. Overhauser shift diminishes
resonantly when the RF-radiation of certain frequencies was applied
to the sample. This effect served as an indirect, yet powerful
method for nuclear magnetic resonance detection: NMR quadrupole
splitting of $^{75}$As nuclei was clearly resolved. Theoretical
calculations performed describe well these experimental findings.

\end{abstract}
 \maketitle

Extensive studies of nuclear spin physics in various semiconductor
heterostructures have been performed in the past several
decades~\cite{Dobers1, Dobers2, Barrett, Tycko:1995, Khandelwal,
Kuzma:1998, Kukushkin:1999, Kronmuller:1999, Melinte, Smet:2002,
Hashimoto:2002, Stern:2004, Kumada:2005, Kumada:2007, Tiemann:2012,
Stern:2012, Tiemann:2014, Friess:2014, Rhone:2015}. Such keen
interest was brought about in view of both applied and fundamental
significance of the topic. Nuclear spins may be utilized to store
information~\cite{Privman:2002,Smet:2002} in terms of spin-based
electronics~\cite{Datta:1990,Wolf:2001, Awschalom:2002, Zutic:2004,
Bader:2010}, as non-equilibrium spin polarization of lattice nuclei
may have extremely long lifetime~\cite{Dobers2}. On the other hand,
fundamental properties of the two-dimensional conduction electrons
and nuclear spins are interconnected. The effect of huge
longitudinal resistance near certain fractional
fillings~\cite{Kronmuller:1998, Kronmuller:1999} may be mentioned as
one of the brightest examples. Moreover, the ground state spin
polarization of the two-dimensional electron system (2DES) can be
extracted from the Knight shift~\cite{Knight} of the nuclear
magnetic resonance (NMR)~\cite{Barrett, Tycko:1995, Melinte,
Stern:2004, Kumada:2007}. This offers the approach to investigate
various 2DES exotic states including Wigner
crystal~\cite{Tiemann:2014} and $\nu=5/2$
state~\cite{Tiemann:2012,Stern:2012,Friess:2014}.

One of the most fruitful approaches is to access nuclear spins
experimentally through the spins of conduction electrons coupled to
them by the hyperfine interaction. Spin properties of the electrons,
in turn, can be effectively studied with the aid of electron spin
resonance (ESR). One of the earliest adaptations of this principle
for the experiments on GaAs-based heterostructures can be found in
the papers~\cite{Dobers1,Dobers2}. Let us address the idea of
studying nuclear spins through ESR in more details. The actual
magnetic field position of ESR turned out to be dependent on the
spin polarization of the nuclear system. Indeed the spin part of the
Hamiltonian for the single electron in a $[001]$ quantum well can be
expressed as:
\begin{equation}
H=g^*{\mu_B} BS_z +\textbf{I}\hat{A}\textbf{S}
\end{equation}
Here $g^*$ is the bare electron
$g$-factor% effective Lande factor of an electron
, which does not take into account the electron-electron exchange
interaction~\cite{Ando-rev,Shkolnikov2002}, $\mu_B$ is the Bohr
magneton, $B$ is the amplitude of the magnetic field, which is
applied along
%magnetic field oriented along the axis $Oz
$z\parallel [001]$, $\textbf{S}=(S_x,~S_y,~S_z)$ is the spin of an
electron, $\textbf{I}=(I_x,~I_y,~I_z)$ is the total nuclear spin,
and $\hat{A}$ is the hyperfine interaction tensor. Provided the
total nuclear spin polarization is non-zero the electron spin
splitting $\Delta E$ in the structure under study can be presented
as
\begin{equation}
\Delta E=g^*\mu_b
S_z{\left(B+\dfrac{A_{zz}I_z}{g^*\mu_B}\right)}=g^*\mu_b S_z
(B+\Delta B) \label{OV}
\end{equation}
The term $\Delta B={A_{zz}I_z}/({g^*\mu_B})$ represents the
Overhauser shift~\cite{Overhauser} of the ESR actual magnetic field
position. Under typical experimental conditions thermal energy is
several orders of magnitude larger than nuclear spin splitting and
nuclear spins are unpolarized, hence, Overhauser shift equals zero.
When the electron system is in ESR, nuclear spins become partially
polarized as part of the non-equilibrium spin polarization is
transferred from the electrons to the nuclear subsystem via
hyperfine interaction. By adjusting the external magnetic field so
that the resonance conditions for ESR are fulfilled at a rather long
period of time it is possible to significantly polarize the nuclear
subsystem and to achieve large Overhauser
shifts~\cite{Dobers1,Dobers2}. This process is traditionally
referred to as dynamic nuclear polarization.

Now the approach for studying nuclear spin subsystem through ESR
becomes obvious. As the Overhauser shift is proportional to the
total spin of the nuclei, the rate of the nuclear spin relaxation is
exactly the decay rate of this shift, and thus the nuclear spin
lattice relaxation rate can be accessed
experimentally~\cite{Dobers2,Ryzhov2015,OpticalField}. Resonant
depolarization of nuclear spins will result in resonant diminishing
of Overhauser shift allowing the effective NMR
detection~\cite{Dobers1}.

 In the present paper we report ESR studies of nuclear
spins in close proximity to 2DES formed in the AlAs-quantum well.
Such a semiconductor heterostructure boasts several peculiar
properties. First of all, nuclear spin lifetime turned out to be
quite long (several hours). Moreover, in wide $[001]$ quantum wells
(wider than 5~nm) the electrons tend to occupy two in-plane valleys
located at the X-points of the Brillouin zone along $[100]$ and
$[010]$, while in narrow quantum wells the X valley along
 $[001]$ has lower energy (see Ref.~\onlinecite{Shayegan:2006}). These valleys are
characterized by an anisotropic effective mass~\cite{Muravev:2015}:
$m_t=0.2~m_0$ and $m_l=1.1~m_0$, much heavier than in conventional
GaAs heterostructures. As a consequence, the ratio between
characteristic Coulomb energy and Fermi energy is by far larger and
thus the many particle effects are significantly more pronounced in
AlAs quantum wells than in GaAs heterostructures. Finally, the value
of conduction electron g-factor is large $g^*\approx1.98$(see Ref.
~\onlinecite{Dietsche:2005,Shchepetilnikov:2015}), whereas the
g-factor in GaAs heterostructures depends strongly on the parameters
of the structure and the magnetic field~\cite{Shchepetilnikov:2013}
but its absolute value does not exceed $0.44$. As a result, in AlAs
quantum wells electron spin splitting also exceeds the thermal
energy correspondent to the temperature of the experiment,
$T=1.5$~K. This results into large intensity of ESR, hence, allowing
for the accurate measurements.

 The sample under study was a 16~nm AlAs quantum well epitaxially grown
along the $[001]$ direction. The Al concentration in the
Al$_x$Ga$_{1-x}$As barrier layers was equal to $46$\%. The structure
was asymmetrically delta-doped with Si to result in a low
temperature sheet density of $n\approx 1.8\times10^{11}$~cm$^{-2}$.
The electron mobility was equal to $\mu=2\times10^5$~cm$^2$/V~s at the
temperature of 1.5~K. Standard indium contacts to the 2D electron
system were formed in the common Hall bar geometry. Low temperature
characterization of this exact sample can be found in our previous
publication~\cite{Shchepetilnikov:2015}.

The conventional method of ESR detection in 2DES is based on the
sensitivity of the system longitudinal magnetoresistance $R_{xx}(B)$
to the spin resonance in the quantum Hall regime~\cite{Stein:83}.
The ESR is detectable as a sharp peak in $R_{xx}(B)$ magnetic field
dependence at a fixed microwave frequency. We have successfully
applied this approach to carefully investigate the g-factor
anisotropy in GaAs - based
heterostructures~\cite{Nefyodov:2011a,Nefyodov:2011b}.

\begin{figure}[th!]
\centerline{\includegraphics[width=0.69\columnwidth,clip]{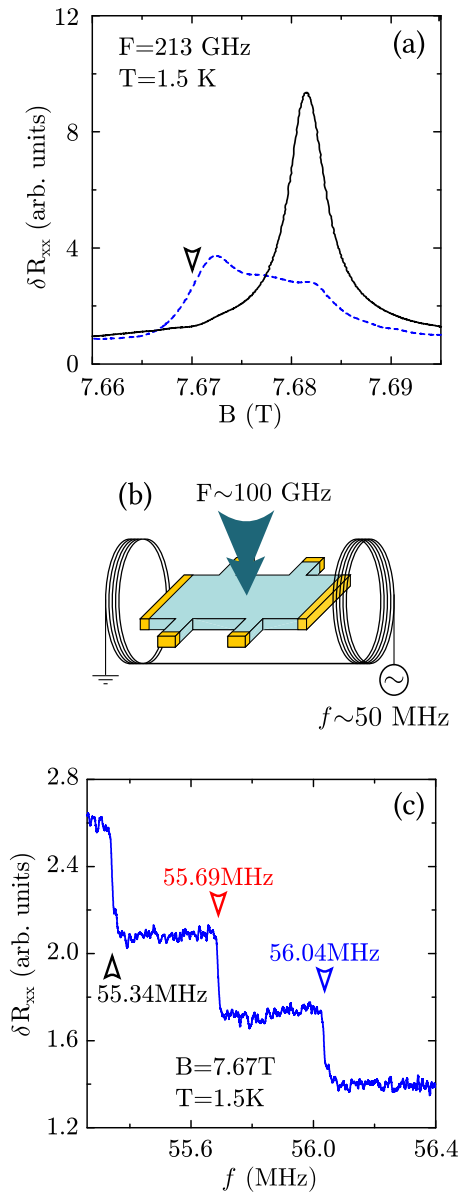}}
\caption{(a) Typical ESR lineshapes observed in the experiment.
Solid line corresponds to the initial ESR, dashed line denotes the
ESR peak after dynamical nuclear polarization. A mark denotes the
fixed value of magnetic field for the NMR experiment. (b) Schematic
of NMR experimental setup. A coil coupled to an RF-source is placed
around the sample. The microwave F$\sim 100$~GHz (via waveguide) and
radio frequency $f\sim 50$~MHz (via coil) radiation can be applied
to the sample. (c) Typical magnetoresistance in the NMR experiment.
The steps correspond to three NMR frequencies of $^{75}$As (nuclear
spin $I=3/2$) split by quadrupole interaction.} \label{NMR_1}
\end{figure}

\begin{figure}[t]
\centerline{\includegraphics[width=0.79\columnwidth,clip]{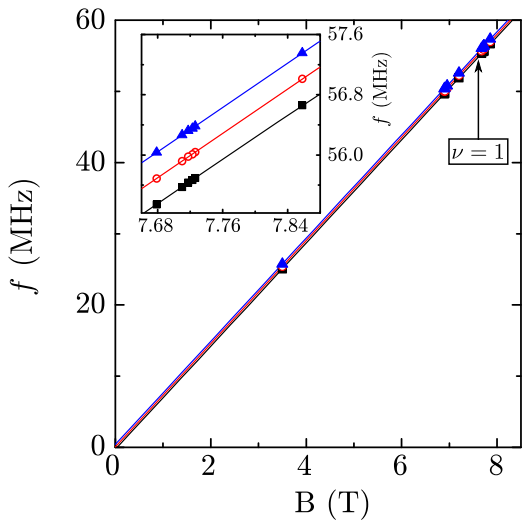}}
\caption{The dependence of three NMR frequencies on the magnetic
field. The inset shows the region of magnetic fields near unity
filling factor in more detail. } \label{NMR_1D}
\end{figure}

An ac probe current of $1~\mu$A at the frequency of $\sim$1~kHz was
applied from source to drain. A lock-in amplifier monitored the
channel resistance $R_{xx}$ through two sense contacts along the
channel. The sample was irradiated by 100\% amplitude modulated
radiation at a frequency of $f_{mod}\sim$30 Hz; microwave power was
delivered through a rectangular oversized waveguide. A number of
microwave radiation sources were utilized: generators with frequency
multipliers coupled to them, backward wave oscillators - so that we
were able to vary the microwave radiation frequency up to 260~GHz.
The power of the microwave radiation injected into the waveguide did
not exceed 10~mW. A second lock-in amplifier, synchronized at
$f_{mod}$ frequency, was connected to the output of the first one
and, thus, measured the variation $\delta R_{xx}$ in the
magnetoresistance, caused by microwave irradiation. Experiments were
carried out at the temperature of $1.5$~K in the magnetic field up
to 10~T.

No traces of dynamic nuclear polarization were observed in our
previous ESR studies~\cite{Shchepetilnikov:2015} of AlAs quantum
well. As the rates of the magnetic field sweeps were high, the
electron spins remained in resonance for only limited periods of
time and, as a consequence, the amount of angular momentum
transferred to the nuclear subsystem was negligible. In order to
form substantial nuclear spin polarization and to achieve large
Overhauser shifts the following procedure was implemented during
present work. The microwave resonance frequency was kept constant.
The magnetic field was ramped to the exact ESR position. The
gradually building up nuclear spin polarization shifted ESR towards
lower fields and the magnetic field was then constantly adjusted so
that the resonant conditions for ESR were fulfilled for a long
period of time. Typically, it took several hours to achieve
substantial Overhauser shifts under our experimental conditions.
Panel (a) of the Fig.~\ref{NMR_1} depicts typical ESR lineshapes
with (dashed line) and without ESR induced nuclear spin polarization
(solid line) measured near unity filling at 1.5 K. The Overhauser
shift of about 10~mT is clearly resolved. We were not able to
achieve Overhauser shifts larger than 40~mT, the value several
orders of magnitude smaller than the one observed in conventional
GaAs-heterostructures~\cite{Kukushkin:1999}. According to the
Eq.~\ref{OV} this effect stems from the relative weakness of
hyperfine interaction in AlAs and large effective electron
$g$-factor. Another important difference was the ESR lineshape in
presence of nuclear spin polarization: in GaAs-based samples ESR
shifts as a whole~\cite{Dobers1}, whereas in our experiments ESR
splits into separate peaks. The right-most peak retains the initial
position, while the left-most one is shifted towards lower magnetic
fields.

 The nuclear origin of the observed shift was proved by NMR
experiments. NMR also allowed us to identify the isotope
participating in the dynamic nuclear polarization. For NMR studies a
coil coupled to an RF-source was placed around the sample, so that
both the microwave $\sim 100$~GHz (via waveguide) and radio
frequency $\sim 50$~MHz (via coil) radiation could be applied to the
sample. The schematic illustration of the experimental setup is
presented in panel (b) of the Fig.~\ref{NMR_1}. The NMR procedure
was as follows. The microwave frequency was kept constant and
Overhauser shifts large enough to split ESR into well-resolved peaks
were achieved (see panel (a) of the Fig.~\ref{NMR_1}). Then the
magnetic field was fixed at the position of the lower (in the
magnetic field) side of the ESR peak experiencing Overhauser shift
(the correspondent magnetic field is indicated by the arrow in panel
(a) of the Fig.~\ref{NMR_1}) and then the sweep of the RF-radiation
frequency was performed. Near NMR nuclear spins were depolarized and
the Overhauser shift resonantly diminished, so that the left-most
(on the magnetic field scale) peak moved towards higher fields, as a
result, the signal measured by the second lock-in amplifier dropped.
Typical signal of the second lock-in amplifier during the
RF-radiation frequency sweep is shown in panel (c) of the
Fig.~\ref{NMR_1}. Three resonant frequencies resolved are indicated
in the panel, the middle one corresponds to the NMR frequency of
$^{75}$As isotope. No traces of other isotopes including Al and Ga
at the corresponding frequencies were found.

The results of the NMR experiments performed near different magnetic
fields and filling factors ($\nu=1$ and $\nu=2$) are presented in
the Fig.~\ref{NMR_1D}. Three resonant frequencies were resolved at
each magnetic field studied. These frequencies are linear in the
magnetic field, whereas differences between the central frequency
and two auxiliary ones are equal to each other and remain constant
$\delta f=0.34$~MHz independent of the magnetic field. We attribute
this splitting to the quadrupole interaction of arsenic nuclei. The
nuclear spin of $^{75}$As equals $3/2$, so the magnetic field splits
the nuclear spin level into four sublevels with different nuclear
momentum projections. Quadrupole interaction modifies the energy
separations between these sublevels and during NMR exactly three
transitions with different energies become possible resulting into
three resonant frequencies of nuclear resonance. Please note, that
the quadrupole splitting of the nuclei spins does not depend on the
magnetic field. The strength of the splitting can be estimated using
the lattice mismatch between the substrate and the quantum
well~\cite{Dzhioev2007}, which is about 0.2\%. The simplest model
calculations~\cite{Suppl} yield the quadrupole splitting 0.3~MHz
which is in a good agreement with the experimental value.

\begin{figure}[t]
\centerline{\includegraphics[width=0.65\columnwidth,clip]{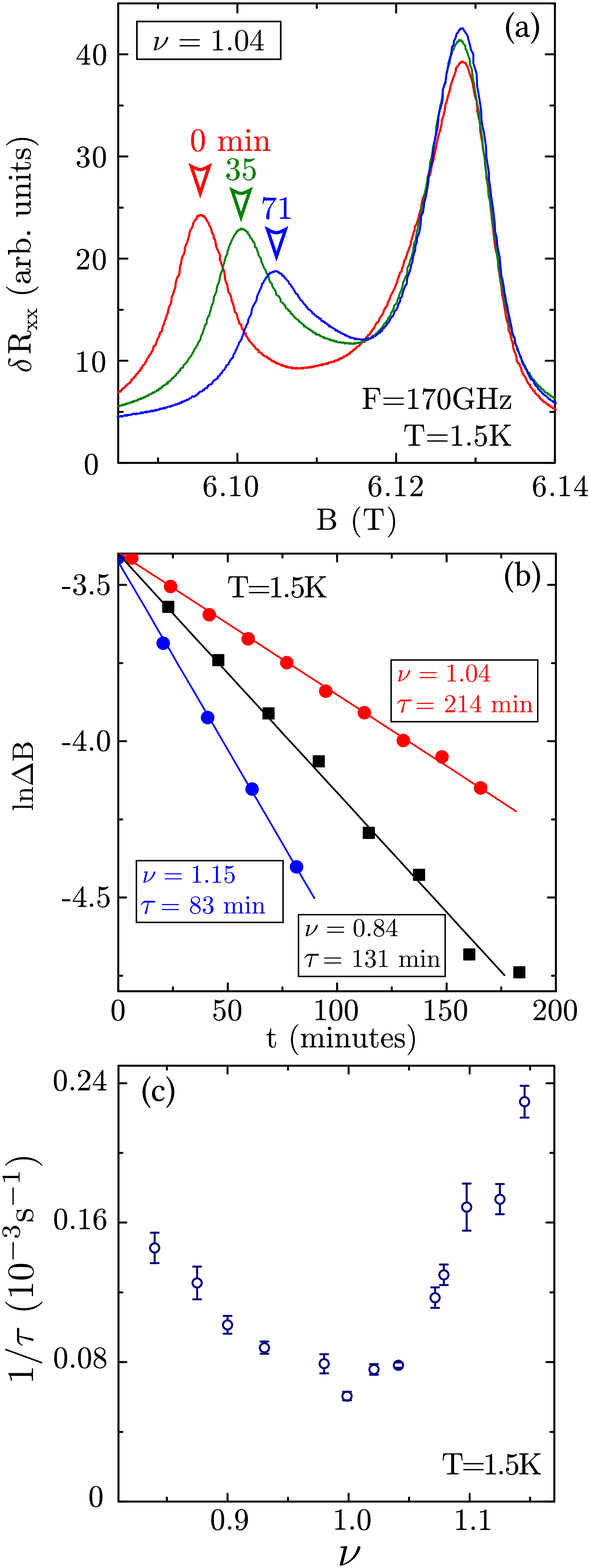}}
\caption{(a) Three consecutive ESR peaks recorded at different
moments of time during the process of nuclear spin relaxation near
filling factor $\nu=1.04$. Elapsed time is indicated near each peak.
Initial nuclear spin polarization was achieved with the aid of ESR
at a fixed frequency of 170~GHz. (b) The dependencies of the
Overhauser shift on time during nuclear spin relaxation near various
filling factors $\nu=0.84,~1.04,~1.15$. The data is fitted with
straight lines to extract the characteristic decay time denoted in
the panel. (c) The nuclear spin relaxation rate measured near
different fillings. The temperature was fixed at $1.5$~K throughout
the experiment. } \label{NMR_2}
\end{figure}

In order to measure the nuclear spin-lattice relaxation time,
$\tau$, non-equilibrium spin polarization of nuclei was first
induced with the aid of ESR  in the vicinity of unity filling factor
at a fixed frequency of 170~GHz. Note, that this whole set of
experiments was performed after another cooldown of the sample and
thus electron sheet density was slightly different $n\approx
1.5\times 10^{11}$~cm$^{-2}$. The decay of this polarization with
time was measured near different fillings as follows. The ESR peak
was measured to probe the Overhauser shift, then the magnetic field
was quickly ramped to the position correspondent to a filling factor
of interest, where nuclear spins relaxed for a relatively long
period of time. Afterwards the magnetic field was ramped back to the
initial position to probe Overhauser shift once again. These steps
were repeatedly performed several times. Three consecutive ESR peaks
recorded during this procedure for filling factor $\nu=1.04$ are
presented in the panel (a) of the Fig.~\ref{NMR_2}. The slow
relaxation of the Overhauser shift with time can be clearly seen.
Typical dependencies of the Overhauser shift on time are plotted in
panel (b) of the Fig.~\ref{NMR_2} for three different fillings. All
of the measured dependencies were exponential so that the decay time
could be extracted.

The nuclear spin relaxation rates measured near different filling
factors around $\nu=1$ are presented in panel (c) of the
Fig.~\ref{NMR_2}. The relaxation rate is clearly dependent on the
state of the electron system and roughly follows the dependence of
DOS at the Fermi level on the filling factor. This fact suggests
that the main relaxation channel is based on the hyperfine
interaction between electron and nuclear spins. However the spin
relaxation mechanisms related to the
  scattering assisted~\cite{Dobers2} and phonon assisted~\cite{Kim1994}
  spin flips
  are suppressed by the large spin splitting, low temperature and
  high quality of the structure. This indicates that some other
  mechanism might be dominant in this case~\cite{Suppl}, as e.g. the hyperfine
  mediated nuclear spin diffusion~\cite{Latta2011}, which does not involve real
  electron spin transitions and allow for the fast long-range nuclear
  spin transport.

Typical nuclear relaxation times measured around unity fillings are
about $200$ minutes long and are of the order of magnitude larger
than in conventional GaAs heterostructures~\cite{Dobers2}. This
experimental finding indicates the relatively small strength of
hyperfine interaction in AlAs.

In order to calculate hyperfine interaction constant microscopically
we have adopted the method outlined in
Ref.~\onlinecite{Gryncharova}. We have decomposed electron Bloch
function in the vicinity of the nuclei into the hydrogen-like  $s$,
$p$, $d$ contributions~\cite{Fischer2008,Chekhovich2013} with the
coefficients determined from the DFT calculation~\cite{Suppl}. The
calculation shows that the hyperfine interaction between the
electron spin and the spin of arsenic nuclei in AlAs is suppressed
by a factor of $\approx2$ as compared to GaAs.
%constitutes approximately 32\% of that in GaAs,
The maximum Overhauser shift is estimated to be 250~mT, while for
GaAs it is known to be 3.53~T~\cite{Urbaszek}. Note that the degree
of nuclear spin polarization in our experiment is about 16\%, which
is somewhat larger than in Ref.~\onlinecite{Dobers2}.

The main reasons for the drastic difference of Overhauser fields in
AlAs and GaAs are the reduced contribution of $s$ shells to the
electron Bloch function in the X point of AlAs Brillouin zone and
the large effective $g$-factor. The main contribution to the
hyperfine interaction is given by the arsenic nuclei, while the
contribution from aluminum is negligible, in full agreement with the
experiment. This is caused by the $p$ type symmetry of Al Bloch
function and the shift of electron density from cation to anion. The
hyperfine interaction tensor is found to be anisotropic due to the
reduced symmetry of the $X$ valleys in AlAs. Microscopically the
main contribution to the hyperfine coupling constant is related with
the $s$ shell of As atom and the anisotropy of hyperfine interaction
is determined by the contribution of its
 $d$ shell~\cite{Suppl}.

% because of smaller contribution of $s$ and $s^*$ shells to the wave function in the $X$
% valley of Brillouin zone. The hyperfine interaction with the
% %aluminum nuclei turns out to be less than 1\% of this value
% cation does not exceed 1\% of this value due to
% %is about 80 times weaker and does not allow to observe the Al
% %nuclear spin polarization in the experiment. Microscopically this
% %strong suppression of hyperfine interaction is related to
% the small mass of the Al nuclei~\cite{Herman1963}. The hyperfine
% interaction for both species is found to be nearly isotropic despite
% the reduced symmetry of the $X$ valleys.

% Neglecting the aluminum contribution the maximum Overhauser field in
% limit of complete polarization is estimated to be 130~mT. Therefore
% the nuclear spin polarization achieved in the experiment can be
% estimated to be about 30\%.

To conclude, the Overhauser shift of the two-dimensional conduction
electron spin resonance was studied in an asymmetrically doped {16
nm} AlAs quantum well grown in the $[001]$-direction.
Non-equilibrium nuclear spin polarization was created while the
electron system was in ESR, as part of the magnetic momentum was
transferred to the nuclear subsystem. The NMR experiments revealed
the nuclear isotope participating in dynamic nuclear polarization to
be $^{75}$As. The NMR was detected by the resonant reduction of
Overhauser shift after the RF-radiation of certain frequencies was
applied to the sample. The quadrupole splitting of $^{75}$As was
observed in the NMR experiment. The nuclear spin-lattice relaxation
rate was measured by the decay of Overhauser shift over time near
different filling factors. The dependence of the relaxation rate on
the state of the electron system suggests that the relaxation
mechanism is based on the hyperfine interaction between nuclear and
electron spins. The maximum Overhauser shifts achieved in the
experiments were substantially smaller than in conventional GaAs
quantum wells and heterojunctions. This fact suggests that hyperfine
interaction is weak in AlAs. The microscopical calculations were
conducted to prove this statement.

We gratefully acknowledge the financial support of the ESR
measurements from Russian Science Foundation (Grant No. 14-12-00693)
and cryogenic NMR setup development from Russian Foundation for
Basic Research (Grant No. 16-32-00399). D. S. S. is grateful to M.
O. Nestoklon and M. M. Glazov for fruitful discussions and
acknowledges partial support from the Russian Science Foundation
(No. 14-12-501067), RF President Grant No. SP-643.2015.5, Dynasty
Foundation, and St.-Petersburg Government Grant. L.T., C.R., W.D.
and W.W. acknowledge the financial support from Swiss National
Foundation ('Quantum Science and Technology').

\onecolumngrid
\vspace{\columnsep}
\begin{center}
\newpage{\large\bf {Supplemental Material to\\
 ``Nuclear magnetic resonance and nuclear spin relaxation in AlAs quantum well probed by ESR''}
}
\end{center}
\vspace{\columnsep}
\twocolumngrid

\renewcommand{\cite}[1]{{[}\onlinecite{#1}{]}}

\renewcommand{\thepage}{S\arabic{page}}
\renewcommand{\theequation}{S\arabic{equation}}
\renewcommand{\thefigure}{S\arabic{figure}}
\renewcommand{\thetable}{S\arabic{table}}
\renewcommand{\bibnumfmt}[1]{[S#1]}
\renewcommand{\citenumfont}[1]{S#1}
\setcounter{page}{1}
\setcounter{section}{0}
\setcounter{equation}{0}
\setcounter{figure}{0}

\section{S1. Quadrupole splitting estimation}

In GaAs/AlAs heterostructures there is a lattice mismatch $\varepsilon\sim0.2\%$ at low temperatures, which can be used to roughly estimate the strain in the QW. The strain produces build-in electric fields in the structure. The nuclear quadrupole moments interact with the electric field gradients, which leads to the quadrupole splitting of the nuclear spin states, as described by the Hamiltonian
\begin{equation}
  \mathcal{H}_Q=\frac{Q}{2}\left(I_z^2-\frac{I(I+1)}{3}\right).
\end{equation}
In the simplest model~\cite{Dzhioev2007_supp}
the quadrupole splitting is determined by
\begin{equation}
  Q=\frac{3e\mathcal{Q}S\varepsilon}{2I(2I-1)},
\end{equation}
where $\mathcal{Q}$ is the nuclear quadrupole moment and $S$ relates the elastic strains to the electric field gradients. For the arsenic nuclei $\mathcal{Q}=3.1\cdot10^{-25}$~cm$^2$ and one can use $S=13.2\cdot10^{15}$~esu/cm$^3$ measured in GaAs for arsenic nuclei~\cite{Sundfors1976_supp}. Estimation with these values gives the splitting between nuclear spin resonances equal to $Q\approx0.3$~MHz in surprisingly good agreement with the experimental value.

\section{S2. Microscopic calculation of hyperfine interaction}

The hyperfine interaction Hamiltonian has the form~\cite{Abragam_supp}
\begin{equation}
 \hat{\mathcal H}_{\rm{hf}}=2\mu_B\mu_I{\bm I}\left[\frac{8\pi}{3}\hat{\bm s}\delta(\bm r)+\frac{\hat{\bm l}}{r^3}-\frac{\hat{\bm s}}{r^3}+3\frac{{\bm r}\left(\hat{\bm s}\cdot{\bm r}\right)}{r^5}\right],
\label{eq:Ham}
\end{equation}
where $\mu_B$ is the Bohr magneton, $\mu_I$ is the nuclear magnetic moment, $\bm I$ is the nuclear spin, $\hat{\bm l}=-i\left[\bm r\times\bm\nabla\right]$ is the angular momentum operator, and $\bm s$ is the electron spin. The origin of the coordinate frame is chosen at the nucleus. The first term in Eq.~\eqref{eq:Ham} describes contact interaction, while the others stand for the dipole-dipole interaction.

To calculate the hyperfine coupling constants we apply the method introduced in Ref.~\onlinecite{Gryncharova_supp}
and developed in Refs.~\onlinecite{Fischer2008_supp,Chekhovich2013_supp}.
The bulk AlAs is described by the $\mbox{T}_d$ point symmetry group with the minimum of the conduction band in the vicinity of X point of the Brillouin zone~\cite{vurgaftman02_supp}. The local symmetry in this valley is reduced and the symmetry group of the $K_X$ Bloch wave vector is D$_{2d}$.

The irreducible representation of the electron wave function depends on the choice of the central point. Provided it is placed at the As atom the corresponding representation is X$_6$~\cite{Birman_supp}. However as soon as the symmetry center is placed at the Al atom in the same elementary cell the representation is changed to X$_7$. Indeed every new symmetry operation can be presented as the old point symmetry operation plus the translation by the vector $\tau$. In the $K_X$ point of the Brillouin zone such translation is equivalent to the multiplication by the factor $\e^{\i K_X\tau}=\pm1$. This factor belongs to $X_4$ representation of $D_{2d}$ group. Therefore when the central point is chosen at Al atom the wave function transforms according to $X_6\otimes X_4=X_7$ representation~\cite{Birman_supp}.

The tight-binding calculations show that for the central point chosen at As atom the dominant contribution to the orbital part of the wave function belongs to X$_1$ representation~\cite{Nestoklon_supp}. To be specific we consider the electron wave function in the X valley oriented along $z$ direction; the wave functions in the two other valleys can be obtained by rotation of the coordinate frame. We decompose the orbital part of the wave function in the vicinity of each nucleus into the $s$, $p$ and $d$ shells as
\begin{subequations}
  \label{eq:psi}
  \begin{equation}
    \Psi_{\rm As}=\alpha_SS(\theta,\phi)R_s(r)+\alpha_DD_{z^2}(\theta,\phi)R_d(r),
  \end{equation}
  \begin{equation}
    \Psi_{\rm Al}=\alpha_PP_z(\theta,\phi)R_p(r)+\alpha_TD_{xy}(\theta,\phi)R_d(r).
  \end{equation}
\end{subequations}
Here $\alpha_l$ ($l=S,P,T,D$) are the coefficients, the functions $R_{s,p,d}(r)$ are the radial parts of the corresponding atomic orbitals, and the angular dependencies are described by the tesseral harmonics $S$ for $s$ shell, $P_x$, $P_y$, $P_z$ for $p$ orbitals and $D_{xy}$, $D_{yz}$, $D_{xz}$, $D_{x^2-y^2}$, $D_{z^2}$ for $d$ orbitals~\cite{Varshalovich_supp}. Note that hereafter we neglect the electron Bloch wave vector~\cite{Gryncharova_supp} as well as the corrections related to the size quantization.

Calculation of the hyperfine interaction, Eq.~\eqref{eq:Ham}, for the functions in the form of Eq.~\eqref{eq:psi} yields
\begin{equation}
  H_{hf}=A^\perp(I_xs_x+I_ys_y)+A^\parallel I_zs_z,
\end{equation}
where
\begin{equation}
  A^i=\frac{4}{3}\mu_B\mu_IR_s^2(0)\sum_l\left|\alpha_l\right|^2C_l^iM_l.
\label{eq:A}
\end{equation}
Here $i=\perp,\parallel$ and the coefficients $C_l^i$
% Here $i=(\perp,\parallel)$, $l=(S,P,T,D)$ and the coefficients $C_l^i$
%are of the order of unity and
are given in Tab.~\ref{tab:C}. The numbers $M_l$ describe the relative strengths of the corresponding contributions to the hyperfine interaction. By definition $M_S=1$, while for $l=P$ and $l=T,D$
\begin{equation}
  M_l=\frac{1}{R_s^2(0)}\int\limits_0^\infty\frac{R_{p,d}^2(r)}{r}\d r,
\end{equation}
respectively, and we have disregarded the crosscorrelations between different $l$.
% and the small corrections to the orbital functions due to size quantization and the electron Bloch wave vector.

%\addDima{Below we describe consequently calculation of the coefficients $C_l^i$, $M_l$ and $\left|\alpha_l\right|^2$ in Eq.~\eqref{eq:A}.}

\begin{table}
\caption{The parameters $C_l^i$ of the hyperfine interaction constants, Eq.~\eqref{eq:A}.}
\label{tab:C}
\begin{center}
%\begin{minipage}{0.25\textwidth}
\begin{ruledtabular}
\begin{tabular}{ccccc}
% \begin{tabular}{c|c|c|c|c}
$C_l^i$     & $l=S$ & $l=P$     & $l=T$    & $l=D$\\
 \hline
$i=\perp$     & 1 & -3/5     & 3/7 & -3/7 \\
\hline
$i=\parallel$ & 1 & 6/5 & -6/7 & 6/7
\end{tabular}
\end{ruledtabular}
%\end{minipage}
\end{center}
\end{table}

Calculation of $M_l$ can be explicitly done for the particular model
of radial functions. We consider the hydrogen like 
functions and the results of our calculations are
presented in Tab.~\ref{tab:M}. The orbital exponents for these functions were
calculated in Ref.~\onlinecite{Clementi_supp} using the self-consistent-field
function.
We note that 
one can also use Slatter functions with the parameters determined
%the parameters of these function can be also calculated using the
by
 Slater
rules~\cite{Slater_supp} or from the \textit{ab-initio}
calculations~\cite{Benchamekh_supp}.

\begin{table}
\caption{The parameters $M_l$ of the hyperfine interaction constants, Eq.~\eqref{eq:A}, calculated using the hydrogen-like wave functions.}
\label{tab:M}
\begin{center}
%\begin{minipage}{0.25\textwidth}
\begin{ruledtabular}
\begin{tabular}{cccc}
          &  $M_S$ &  $M_P$     & $M_T=M_D$ \\
 \hline
Al     & 1 & 0.08024 &  0.01605 \\
 \hline
Ga     & 1  & 0.05671 & 0.3849 \\
\hline
As & 1 & 0.04815  & 0.2898
\end{tabular}
% \begin{tabular}{cccc}
%           & Al & Ga     & As \\
%  \hline
% $M_S$     & 1 & 1 & 1 \\
%  \hline
% $M_P$     & 0.08024 & 0.05687 & 0.04815 \\
% \hline
% $M_T=M_D$ & 0.01605 & 0.3849  & 0.2898
% \end{tabular}
\end{ruledtabular}
%\end{minipage}
\end{center}
\end{table}

Finally the probabilities of the atomic shell occupations, $\left|\alpha_l\right|^2$ multiplied by the probability to occupy the particular atom, were calculated in the WIEN2k package~\cite{Nestoklon_supp,WIEN2k_supp} using modified
Becke-Johnson (mBJ) exchange-correlation potential~\cite{mBJ_supp}. The results are presented
% The coefficients $|\alpha_l|^2$ multiplied by renormalized with the probabilities for
% electron to be at the given sublattice are given
in Tab.~\ref{tab:P}. One can note that these values somewhat differ
from the accepted ones~\cite{Boguslawski_supp}, but the main features are
generally the same.

\begin{table}
\caption{The contributions of atomic shells to the electron density, $|\alpha_l|^2$ multiplied by the probabilities to be in the given sublattice.}
\label{tab:P}
\begin{center}
\begin{ruledtabular}
\begin{tabular}{cccccc}
        &  & $S$ & $P$ & $T$ & $D$ \\
 \hline
GaAs & Ga     & 0.507 &  &  & \\
% \hline
 & As     & 0.493 &  &  & \\
 \hline
AlAs & Al     &  & 0.263 & 0.087 & \\
% \hline
 & As     & 0.252 &  &  & 0.398
\end{tabular}
\end{ruledtabular}
%\end{minipage}
\end{center}
\end{table}

The hyperfine interaction constants can now be calculated after
Eq.~\eqref{eq:A} and the results are presented in Tab.~\ref{tab:A}.
The hyperfine constants are given in the arbitrary units, because the manybody effects considerably modify electron
wavefunction in the vicinity of the nuclei and the outlined approach allows only for the reliable calculation of the hyperfine coupling constants relative values~\cite{Chekhovich2013_supp,Vidal2016_supp}. The hyperfine constants given in Tab.~\ref{tab:A} in the arbitrary units allow one to estimate the absolute values of the coupling constants comparing the results obtained for AlAs and GaAs conduction bands. 
Since the conduction band minimum in GaAs is
formed only by $s$ orbitals the Eqs.~\eqref{eq:Ham}---\eqref{eq:A} the experiment 
can be equally applied to the $\Gamma$ point of GaAs Brillouin zone.
Note that in the considered AlAs QW the electrons occupy the in-plane X valleys, therefore the experimentally observed $A_{zz}$ in the notations of the main text is given by $A^\perp$.
Assuming that the maximum Overhauser field for completely polarized
arsenic nuclei is 2.76~T ($A=47~\mu$eV)~\cite{Paget1977_supp,Urbaszek_supp}
we obtain the maximum theoretical Overhauser field in AlAs 250~mT.

\begin{table}
\caption{The hyperfine coupling constants calculated after Eq.~\eqref{eq:A}. The values are given in the arbitrary units.}
\label{tab:A}
\begin{center}
\begin{ruledtabular}
\begin{tabular}{cccc}
        &  & cation & anion \\
 \hline
GaAs & $A^\perp=A^\parallel$ & 3.12 & 3.97 \\
\hline
AlAs & $A^\perp$ & -0.03 & 1.63 \\
% \hline
 & $A^\parallel$  & 0.07 & 2.82
\end{tabular}
\end{ruledtabular}
%\end{minipage}
\end{center}
\end{table}

The contribution to the Overhauser field from the Al nuclei is very small because this contribution is related mainly to the $p$ shell, which is characterized by small $M_P$. The corresponding hyperfine interaction is strongly anisotropic, as described by $C_P^\parallel=-2C_P^\perp$. For the Al nuclei $A^\perp=-0.4~\mu$eV and $A^\parallel=0.8~\mu$eV. By contrast for the arsenic atom the considerable fraction of the electron density is in the $s$ shell and gives rise to the quite pronounced hyperfine interaction, $A^\parallel=33~\mu$eV. The $d$ shell of the arsenic also contributes to hyperfine coupling, and induces the anisotropy of hyperfine interaction in AlAs, which results into $A^\perp=19~\mu$eV.

\section{S3. Discussion of nuclear spin relaxation mechanism}

Nuclear spin relaxation rate strongly depends on the Landau level
filling factor, which evidences the hyperfine mediated mechanism of
spin relaxation. However the electron-nuclear spin flip process is
accompanied by the change of electron Zeeman energy. The
electron-electron exchange interaction
%between electrons in the quantum Hall regime
 results into the enhancement of the splitting between spin
sublevels~\cite{Ando-rev_supp}. For the Landau level filling factor
$\nu=1$ ($B=6.36$~T) the enchanced $g$-factor is $g^{**}\approx
9$~\cite{Papadakis1999_supp} and the corresponding spin splitting is
$\Delta E=3.3$~meV. On the other hand the temperature expressed in
energy units is $k_BT=0.13$~meV and the Landau level broadening is
$\Gamma=e\hbar\sqrt{2B/(\pi\mu m_t m_l)}=0.12$~meV~\cite{Weiss1987_supp}.
Therefore the phonon assisted~\cite{Kim1994_supp} and scattering
 assisted~\cite{Dobers2_supp}
electron nuclear spin flip processes are suppressed respectively by
the factors $\exp(-\Delta E/k_b T)$ and $\exp(-\Delta E/\Gamma)$ of
the order of $10^{-12}\div 10^{-13}$. In the similar experiments in
GaAs~\cite{Dobers2_supp}
the suppression is much weaker $\exp(-\Delta E/\Gamma)\sim 10^{-5}$
mainly because of the smaller effective mass. In the same time the
nuclear spin relaxation rate in this structure is only 12 times
faster than in AlAs structure under study. This suggests that the
nuclear spin relaxation can be caused by the processes that do not
involve real electron spin transitions.

The nuclear spin diffusion is caused by the dipole-dipole
interaction between nuclear spins~\cite{Abragam_supp}. The strength of
this interaction is very small $H_{dd}\sim 10^{-11}$~meV, while the
Knight field experienced by the nuclei at $\nu=1$ can be estimated
as $K\sim A_{zz}\Omega n/l\sim10^{-7}$~meV, where
$\Omega=0.045$~nm$^3$ is the elementary cell volume and $l=16$~nm is
the QW width. The Knight field acts as an additional magnetic field,
therefore the gradient of the Knight field along the growth axis
related to the electron envelope wave function blocks nuclear spin
diffusion. The energy barrier for the two nearest nuclei spins
flip-flop process is of the order of $\Delta K\sim
Ka_0/l\approx4\cdot10^{-9}$~meV, where $a_0$ is the GaAs lattice
constant. As a possible mechanism of nuclear spin relaxation we
propose nuclear spin diffusion by electron assisted RKKY
interaction~\cite{Latta2011_supp}. This process does not involve real electron spin
transitions, and the energy difference between initial and final
states can be easily compensated by the electron scattering because
$\Gamma\gg\Delta K$. However the microscopic theory of this effect
is beyond the scope of this paper and will be reported elsewhere. We
note that the other mechanisms related to the spin-orbit
coupling~\cite{Hashimoto_supp} or collective excitations~\cite{Cote_supp}
might also contribute to nuclear spin relaxation.

\end{document}